\documentclass[12pt]{article}
\usepackage{ascmac}
\usepackage{ulem}
\usepackage{latexsym}
\usepackage[dvipdfm]{graphicx}
\usepackage{amsmath}
\usepackage[top=3.5cm,bottom=3.5cm,left=2cm,right=2cm]{geometry}

\newcommand{\al}[1]{\begin{align}#1\end{align}}

\newcommand{\GeV}{\ensuremath{\,\text{GeV} }}

\newcommand{\nn}{\nonumber\\}

\newcommand{\h}{\hspace}
\newcommand{\be}{\begin{equation}}
\newcommand{\e}{\end{equation}}
\begin{document}

\title{
\vbox{
\baselineskip 14pt
\hfill \hbox{\normalsize KUNS-2495}
} \vskip 1cm
\bf \Large Evidence of the Big Fix\vskip 0.5cm
}
\author{
Yuta~Hamada\thanks{E-mail:  \tt hamada@gauge.scphys.kyoto-u.ac.jp},
Hikaru~Kawai\thanks{E-mail: \tt hkawai@gauge.scphys.kyoto-u.ac.jp}, and 
Kiyoharu~Kawana\thanks{E-mail: \tt kiyokawa@gauge.scphys.kyoto-u.ac.jp}
\bigskip\\
\it \normalsize
 Department of Physics, Kyoto University, Kyoto 606-8502, Japan\\
\smallskip
}
\date{\today}

\maketitle

\abstract{\noindent\normalsize
}We give an evidence of the Big Fix.  The theory of wormholes and multiverse suggests that the parameters of the Standard Model are fixed in such a way that the total entropy at the late stage of the universe is maximized, which we call the maximum entropy principle. In this paper, we discuss how it  can be confirmed by the experimental data, and we show that it is indeed true for the Higgs vacuum expectation value $v_{h}$.  We assume that the baryon number is produced by the sphaleron process, and that  the current quark masses, the gauge couplings and the Higgs self coupling are fixed when we vary $v_{h}$. It turns out that the existence of the atomic nuclei plays a crucial role to maximize the entropy. This is reminiscent of  the anthropic principle, however it is required by the fundamental low in our case.  
\\
\\{\it{Keywards}}: Quantum gravity, maximum entropy principle, sphaleron, baryon decay, multiverse, wormhole. 
\newpage

\section{Introduction}
There are many open questions that the Standard Model can not answer. The Cosmological Constant Problem is the biggest one. The naive expectation value of the vacuum energy is given by 
\be \rho\simeq M^{4}_{pl}\simeq10^{72}\text{GeV}^{4}, \e
if we assume that the cutoff scale exists around the Planck scale. However, from cosmological observations, we know that the vacuum energy density of our universe is 
\be \rho_{\Lambda}\simeq10^{-48}\text{GeV}^{4}.\e
The above mismatch between Eq.(1) and Eq.(2) is the Cosmological Constant Problem. There are other important  problems of fine tuning in the Standard Model(SM): the Higgs mass, the Yukawa couplings, the QCD $\theta$ parameter  $\cdots$ and so on. The ordinary approach to these problems is to expect the emergence of a new physics at some high energy scale. For example, the supersymmetry is one of the  candidates to answer the Higgs mass hierarchy. However, the recent observations of the Higgs mass~\cite{Chatrchyan:2012ufa,Aad:2012tfa}, and the related analyses \cite{quartic} show the possibility that  that there is no new physics between the weak and the Planck scale\footnote{The observed Higgs mass may indicate that Higgs is the inflaton~\cite{inflation}.}. If it is the case, we must answer the above problems by the Planck scale physics in the framework of the quantum gravity  or string theory\footnote{The other possibilities are to introduce new physical principles such as the multiple point principle \cite{Froggatt:1995rt,Froggatt:2001pa,Nielsen:2012pu},the classical conformality \cite{Meissner:2006zh}, hidden duality \cite{Kawamura:2013kua} and the asymptotic safety\cite{Shaposhnikov:2009pv}. There is also an interesting argument \cite{Alves:2013lya} which might be related to the maximum entropy principle.}.
\\

One possibility along this direction is to consider the dynamics of wormholes and multiverse\cite{Coleman:1988tj}. Recently, there has been a development \cite{Kawai:2011rj,Kawai:2011qb,Kawai:2013wwa} in which the wave function of the multiverse is constructed based on the Lorentzian path integral. As a result, assuming that we live in the $S^{3}$ universe, it is found that the parameters of the SM are fixed in such a way that the total energy of the universe at the late stage $E_{tot}$ is maximized, see \ref{Multiverse} for the details. It is called The Big Fix. If the universe  at the late stage is radiation(matter) dominated, it means that the total radiation(matter) is maximized, which we call  the maximum entropy(matter) principle\footnote{We regard the total radiation as the total entropy.}. In order to test this principle, it is sufficient to examine how $E_{tot}$ varies when we change each of the SM parameters from the experimental values. If we find that $E_{tot}$ decreases under such change, the principle is verified. It seems that this program can be accomplished only if we know the whole history of the universe. However, for some of the SM parameters, we know how they participate the history of the universe.

In particular, we can check the maximum entropy principle for the Higgs vacuum expectation value $v_{h}$. In this paper, we show that this principle is actually satisfied when $v_{h}$ is around $v_{h}=246$GeV. Here we assume that the dark matter(DM) decays much earlier than baryons so that the universe at the late stage becomes dominated by the radiation $S_{rad}$ produced by the decay of baryons. 
\newpage
\noindent Note that $S_{rad}$ is defined by   
\be S_{rad}=a^{4}\times\rho_{rad}(a)\e
where $\rho_{rad}(a)$ is the energy density of the radiation and $a$ is the radius of the universe. 
\\
\\
The overview of how $S_{rad}$ is produced in the universe is as follows: 
\begin{itemize}\item {\it{Stage 1}}: The baryon number $N_{B}$ is produced by the sphaleron process if we assume the standard leptogenesis scenario.\\
\item {\it{Stage 2}}: The ratio of neutrons to all nucleons $X_{n}$ is fixed by the following two processes: \\ 
    A) The weak interaction is frozen out.\\
    B) Some of the neutrons are converted to protons through the beta decay before the Big Bang Nucleosynthesis (BBN).\\
\item{\it{Stage 3}}: $S_{rad}$ is created by the decay of baryons. There are two contributions; hydrogens(free protons) and atomic nuclei. \end{itemize}
The quantities that appear in the above stages can be expressed as a function of $v_h$, $y_u$ and $y_d$. In fact, at the stage 1, using the recent numerical analysis of the sphaleron process\cite{D'Onofrio:2012ni}, we can determine $N_{B}$ as a function of $v_{h}$ if we fix the gauge couplings, the Higgs self coupling, and the top quark mass.  At the stage 2, $X_{n}$ is determined by the Fermi constant, the proton neutron mass difference and the electron mass. Therefore, it can be expressed as an explicit function of $v_{h}$ and the difference of the Yukawa couplings $y_{-}=y_{d}-y_{u}$. At the stage 3, for the case of the proton decay, $S_{rad}$ can be expressed in terms of the mass and the life time of protons; $m_{p},\tau_{p}$. On the other hands, for the case of the decay of the helium nuclei, the situation is rather complicated. We must consider the effect that the pion produced by the decay of a nucleon loses its energy through a collision with the other nucleons. Then, in addition to the mass and life time of helium nuclei, we need a quantity that indicates the average radiation energy created by the decay of a helium nucleus, which we express as $(1-2\epsilon)m_{p}$. In principle, these five quantities 
\be m_{p}\h{2.5mm},\h{2.5mm}\tau_{p}\h{2.5mm},\h{2.5mm}m_{He}\h{2.5mm},\h{2.5mm}\tau_{He}\h{2.5mm},\h{2.5mm}\epsilon\e
can be calculated by the QCD once the $\Lambda_{QCD}$ and the current quark masses
\be m_{u,d}=y_{u,d}\cdot \frac{v_{h}}{\sqrt{2}}\e
are given\footnote{One need not worry about the GUT scale dependence because it appears as an overall factor.}. However, since it is technically difficult, we try several phenomenological values for them. Using the above results, we can show that $S_{rad}$ in deed takes the maximum  around the experimental values of $v_{h}$.
\\
\newpage
\noindent This paper is organized as follows. In Section2, we discuss the history of the baryons, and give general differential equations that explain the time evolution of the radiation. In Section3, the qualitative and numerical results are shown. In Section5, we give summary and discussion. In \ref{WKB}, we present the WKB solution of the wave function of the universe. In \ref{Multiverse}, the probabilistic interpretation of the wave function of the universe is discussed. In \ref{app:sphaleron}, we study the sphaleron process. In \ref{app:epsilon}, we estimate $\tau_{He}$ and $\epsilon$. In \ref{app:darkmatter}, we consider the scenario in which the universe at the late stage becomes dominated by the DM.

\section{History of Baryons}\label{sec:baryon history}
We first mention the possible scenarios of the universe at the late stage. Since we do not know the fate of the DM, there are typically three possibilities\footnote{We assume that the Cosmological Constant at the late stage of the universe is fixed to the critical value $\Lambda_{cri}$ so that it balances with the curvature. This is one of the predictions of the maximum entropy principle. See \ref{Multiverse} for the details.}:\\
\\
(I): The DM decays earlier than the  baryons, and the universe becomes dominated by the radiation after the baryon decay.\\
\\
(II): The baryons decay earlier than the DM, and the universe becomes dominated by the radiation after the dark matter decay.\\
\\
(III): The baryons decay, but the DM does not, then the universe is dominated by DM.\\
\\
In this paper, we consider the scenario (I), and study how the baryons of the universe are produced at the early universe, and decay at the late stage of the universe. We give our argument along the stages indicated in the introduction. The consideration about the scenario (III) is presented in \ref{app:darkmatter}. 
\subsection*{Stage 1 : Baryogenesis by the Sphaleron Process}
In this subsection, we study how the baryon number is produced and how it depends on the SM parameters. We assume that it is produced by the sphaleron process. If the decoupling temperature of that process is comparable with the heavy quark masses, we must consider the decrease of the baryon number caused by them. The detailed calculations are presented in \ref{app:sphaleron}.

Using the recent numerical results  \cite{D'Onofrio:2012ni}, we can obtain the decoupling temperature of the sphaleron process 
\begin{align} T_{sph}\simeq\frac{7}{8}T_{c}=140\text{GeV}\times\frac{v_{h}}{246\text{GeV}}\end{align}
\newpage
\begin{figure}
\begin{center}
\includegraphics[width=9cm]{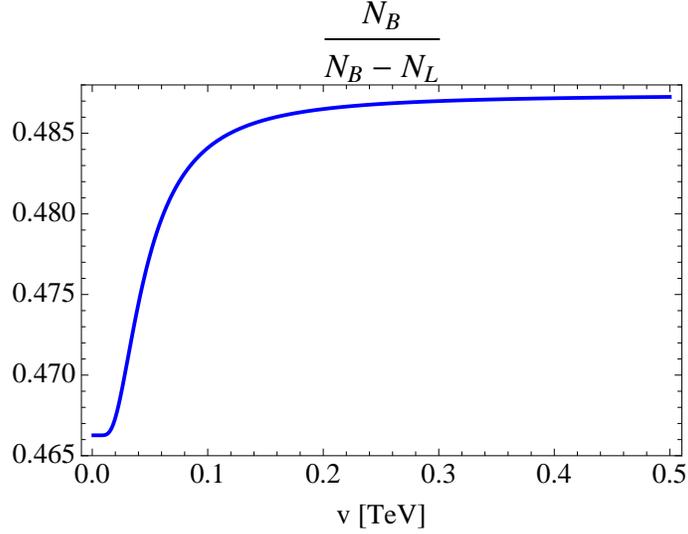}
\caption{The baryon number as a function of $v_{h}$ when the current quark masses, the gauge couplings and the Higgs self coupling are fixed. We can see that $\frac{N_{B}}{N_{B}-N_{L}}$ decreases around $v_{h}=150$GeV corresponding to the decrease of $v_{h}$.}
\label{baryon1}
\end{center}
\end{figure}
\noindent where $T_{c}$ is the critical temperature of the phase transition. Here we fix the gauge couplings and the Higgs self coupling to the observed values, and regard $v_{h}$ as a variable. Furthermore, we can find the temperature dependence of the Higgs expectation value from \cite{D'Onofrio:2012ni}, 
\be v_{h}(T)=v_{h}\sqrt{\frac{T_{c}-T}{T_{c}}}.\label{suppression}\e
This is consistent with the behavior of the Landau theory. Since the particle masses are proportional to $v_{h}(t)$, the suppression factors caused by the quarks and W boson at $T_{sph}$ are given by
\be \frac{m_{i}}{T_{sph}}=\frac{m_{0i}\sqrt{\frac{T_{c}-T_{sph}}{T_{c}}}}{T_{sph}}=\frac{12}{7\sqrt{8}}\times \frac{m_{0i}}{v_{h}}=\frac{3}{7}\times \frac{y_{i}v_{h}}{v_{h}},\label{quarksup}\e
\be  \frac{m_{W}}{T_{sph}}=\frac{m_{0W}\sqrt{\frac{T_{c}-T_{sph}}{T_{c}}}}{T_{sph}}=\frac{12}{7\sqrt{8}}\times \frac{m_{0W}}{v_{h}}=\frac{6}{7\sqrt{8}}\times \frac{g_{2}v_{h}}{v_{h}}\e
where $m_{0i}$ and $m_{0W}$ are the physical masses and $g_{2}$ is the gauge coupling.  
When $v_{h}$ becomes comparable with $m_{i}$, Eq.(\ref{quarksup}) starts to have an effect, and the baryon number decreases. Since $T_{sph}$ is of order $100$GeV, we need not to consider the light quarks. Namely, only the top quark and W boson are crucial to find the maximum of the entropy. In this approximation, we can obtain the baryon number of our universe (see Eq.(\ref{baryon number}) in \ref{app:sphaleron});
\be N_{B}=\frac{N_{B}}{N_{B-L}}N_{B-L}=\frac{2\left(4+g_{b}(m_{W})\right)\cdot\left(5+g_{f}(m_{t})\right)}{24\left(5+g_{f}(m_{t})\right)+g_{b}(m_{W})\left(37+2g_{f}(m_{t})\right)}N_{B-L}\label{bar}\e
where $g_{f}(m_{t})$ and $g_{b}(m_{W})$ are given by 
\be g_{f}(m_{t}):=\frac{3}{2\pi^{2}}\int_{0}^{\infty}dx\frac{x^{2}}{\cosh^{2}\left(\frac{\sqrt{x^{2}+(m_{t}/T)^{2}}}{2}\right)},\nonumber\e
\be g_{b}(m_{W}):=\frac{3}{2\pi^{2}}\int_{0}^{\infty}dx\frac{x^{2}}{\sinh^{2}\left(\frac{\sqrt{x^{2}+(m_{W}/T)^{2}}}{2}\right)}.\e
If we fix the Yukawa couplings and the gauge couplings, there is no $v_{h}$ dependence. Therefore, in this case, the baryon number does not depend on $v_{h}$. On the other hands, if we fix the current quark masses, Eq.(\ref{quarksup}) depends on $v_{h}$.  As we will discuss in the following sections, it is appropriate to fix the current quark masses, the gauge couplings and the Higgs self coupling when we vary $v_{h}$. We plot Eq.(\ref{bar}) as a function of $v_{h}$ in Fig.\ref{baryon1}. Especially, since the coupling constants are fixed, the suppression factor of the W boson does not depend on $v_{h}$. One can see that $N_{B}$ starts to decrease around $v_{h}\simeq150$GeV when we make $v_{h}$ small. This is, of course,  the effect of the top quark mass. In the history of the universe, one might think that the top quark hardly plays an important role. However, the above result suggests that it has an effect in determining the baryon number. We will use Eq.(\ref{bar}) to calculate the radiation.

\subsection*{Stage 2 : Neutron to Baryon Ratio $X_{n}$}
In this section, we discuss how $X_{n}$ is determined at the early universe. Because the total number of helium nuclei is given  by
\be N_{\text{He}}=N_{B}\frac{X_{n}}{2},\e
the determination of $X_{n}$ corresponds to that of $N_{\text{He}}$. We denote the photon temperature by $T$. For T$>>$1MeV,  neutrons and protons are in thermal equilibrium through the six processes:
\be n+\nu\leftrightarrow p+e^{-}\h{4mm}n+e^{+}\leftrightarrow p+\bar{\nu}\h{4mm}n\leftrightarrow p+e^{-}+\bar{\nu},\e
and $X_{n}$ is given by
\be X_{n}=\frac{1}{1+\exp(\frac{Q}{T})}.\e
where $Q:=m_{n}-m_{p}$ is the mass difference between a proton and a neutron. However, at a temperature between 1MeV and 0.1MeV, the reaction rates of the above processes except for the beta decay become smaller than the expansion rate of the universe. This decoupling temperature $T_{dec}$ is determined by
\be \Gamma(p\rightarrow n)=H:=\frac{\dot{a}}{a}\label{decoupling}\e
where $\Gamma(p\rightarrow n)$ is the proton to neutron reaction rate and $H$ is the Hubble expansion rate. For $T>>$1MeV, $\Gamma(p\rightarrow n)$ is given by \cite{Weinberg}
\be\Gamma(p\rightarrow n)
=0.400\text{sec$^{-1}$}\times\left(\frac{T}{1\text{MeV}}\right)^{5}\times\left(\frac{246\text{GeV}}{v_{h}}\right)^{4}, \label{reaction rate}\e

\begin{figure}
\begin{center}
\includegraphics[width=8cm]{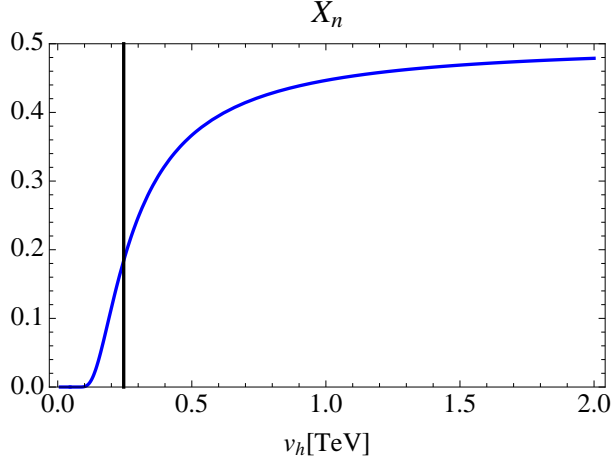}
\label{xngraph}
\caption{The value of Xn as a function of $v_{h}$ when the current quark masses are fixed. We can see that $X_{n}$ depends strongly on $v_{h}$. For example, if $v_{h}\simeq100$GeV, $X_{n}$ is almost zero, and there are no atomic nuclei.}
\end{center}
\end{figure}
\noindent and the expansion rate at the time $t$ is
\be H=\frac{1}{2t}=\frac{1}{2}\sqrt{\frac{2\pi^{2} {\cal{N}}}{45M^{2}_{pl}}}T^{2}\label{eq:expansion},\e
where ${\cal{N}}$ represents the degree of freedoms. By solving Eq.(\ref{decoupling}), we can determine $T_{dec}$ as a function of $v_{h}$ \footnote{Note that, since $T_{dec}$ is an increasing function of $v_{h}$, if $T_{dec}$ becomes comparable with the electron mass,  we must consider the effect of the electron-positron annihilation.}.  
\\

Below $T_{dec}$, $X_{n}$ decreases through the beta decay process, and we obtain
\be X_{n}(t)=\exp(-\tau_{n}^{-1}(t-t_{dec}))\times X_{n}(t_{dec})=\exp(-\tau_{n}^{-1}(t-t_{dec}))\times\frac{1}{1+\exp(\frac{Q}{T_{dec}})}\e  
where $t_{dec}$ is the decoupling time, and $\tau_{n}$ is the neutron life time\footnote{The overall coefficient depends on the current quark masses and the electron mass, which are not important because we fix them in this paper.}
\be \tau_{n}^{-1}=885^{-1}\text{sec}^{-1}\times\left(\frac{246\text{GeV}}{v_{h}}\right)^{4}.\e
When $T$ reaches the BBN temperature $T_{BBN}$\footnote{$T_{BBN}$ depends on the current quark masses and $v_{h}$. However, since its dependences are rather weak, we regard it as a constant in the following discussion.}($\simeq0.1$MeV), neutrons are convert to helium nuclei rapidly. Thus, $X_{n}$ is fixed to the value given by
\be X_{n}=\exp(-\tau_{n}^{-1}(t_{\text{BBN}}-t_{dec}))\times\frac{1}{1+\exp(\frac{Q}{T_{dec}})},\label{Xnvalue}\e
where $t_{BBN}$ is the BBN time. Using Eq.(\ref{eq:expansion}), we can convert $t_{dec}$ and $t_{BBN}$ to $T_{dec}$ and $T_{BBN}$. Fig.\ref{xngraph} shows $X_{n}$ as a function of $v_{h}$ when we fix the current quark masses. One can see that $X_{n}$ depends strongly on $v_{h}$. 

\subsection*{Stage 3 : Baryon Decay and Production of Radiation}
As is discussed in the introduction, the radiation due to the baryon decay comes from not only free protons but also the nucleons in helium nuclei. 
Before considering the realistic case, we first study the simplified case \cite{Kawai:2011qb} where the baryons are all protons. In this case, $S_{rad}$ can be estimated as follows. If we simplify the situation so that protons decay simultaneously, from the energy conservation, we have
\be \frac{N_{B}m_{p}}{a(\tau_{p})^{3}}=\frac{S_{rad}}{a(\tau_{p})^{4}}\label{eq:energy low}\e
where $\tau_{p}$ is the proton life time. From Eq.(\ref{eq:energy low}), we have
\be S_{rad}=N_{B}m_{p}a(\tau_{p}).\label{simpleradiation}\e
Then we use the Friedman equation to express $a(\tau_{p})$ in terms of $\tau_{p}$;
\be \frac{1}{\tau_{p}^{2}}\simeq\frac{1}{M_{pl}^{2}}\frac{N_{B}m_{p}}{a(\tau_{p})^{3}},\e
and we obtain
\be S_{rad}=constant\times(\frac{1}{M_{pl}^{2}})^{\frac{1}{3}}\times (N_{B}m_{p})^{\frac{4}{3}}\tau^{\frac{2}{3}}_{p}.\label{simplerad}\e
This is the qualitative expression of the radiation without atomic nuclei. In \cite{Kawai:2011qb}, it is discussed that the values of the current  quark masses $m_{u},m_{d}$ may be determined by requiring that Eq.(\ref{simplerad}) should be maximized. 
 
Let's start to consider the effect of atomic nuclei. We denote the number of the protons at the time $t$ by $N_{p}(t)$, and that of the helium nuclei by $N_{He}(t)$. They decrease as follows \footnote{The neutrons produced by the decay of helium nuclei are converted to protons through the beta decay immediately compared to the cosmological time.}:
\be \frac{dN_{p}(t)}{dt}=-\tau_{p}^{-1}\cdot N_{p}(t)+3\tau^{-1}_{He}\cdot N_{He}(t)\e
\be  \frac{dN_{He}(t)}{dt}=-\tau_{He}^{-1}\cdot N_{He}(t).\e
The Friedman equation determines the evolution of the scale factor;
\be \left(\frac{\dot{a}}{a}\right)^{2}=\frac{1}{3M_{pl}^{2}}\cdot\left(\frac{M(t)}{a^{3}}+\frac{S_{rad}(t)}{a^{4}}-\frac{M_{pl}^{2}}{a^{2}}+M_{pl}^{2}\h{1mm}\Lambda\right)\label{friedman}\e
where
\be M(t)=m_{p}N_{p}(t)+m_{He}N_{He}(t)
\e 
is the amount of the matter, and $m_{He}$ is the mass of a helium nucleus. In this paper, we express it as
\be m_{He}=2m_{p}+2m_{n}-\Delta\e
where $\Delta=28$MeV is the binding energy. Since what we need is the total radiation $S_{rad}$, we can neglect the curvature term and the cosmological constant term in Eq.(\ref{friedman}). The increase rate of the radiation is given by
\be \frac{dS_{rad}(t)}{dt}=a(t)m_{p}\times\left(\tau_{p}^{-1}\cdot N_{p}(t)+(1-2\epsilon)\cdot\tau_{He}^{-1}N_{He}(t)\right).\label{radiation2}\e
Here we have assumed that while each proton creates the radiation $m_{p}$ when it decays, each helium nucleus creates the radiation $(1-2\epsilon)m_{p}$. The initial values of $N_{p}(t),N_{He}(t)$ are given by
\be N_{p}(0)=N_{B}(1-2X_{n}),\e
\be N_{He}(0)=N_{B}\frac{X_{n}}{2},\e
where we use Eqs.(\ref{bar})(\ref{Xnvalue}) as $X_{n}$ and $N_{B}$. Using the above equations, we can calculate $S_{rad}$ if we give the five quantities 
\be m_{p}\h{2.5mm},\h{2.5mm}\tau_{p}\h{2.5mm},\h{2.5mm}m_{He}\h{2.5mm},\h{2.5mm}\tau_{He}\h{2.5mm},\h{2.5mm}\epsilon.\e
In principle,  these quantities are determined by the QCD. In particular, if we fix the current quark masses, they do not depend on $v_{h}$. Therefore, the $v_{h}$ dependence appears only in $N_{B}$ and $X_{n}$.

\section{The Big Fix of $v_{h}$}\label{sec:big fix}
In this section, we first give qualitative arguments and then give numerical results. As discussed before, we study whether the radiation of the universe becomes maximum around the experimental value $v_{h}=246$GeV.  

\subsection{The Qualitative Argument}\label{subsec:qualitaive}
\begin{figure}[!h]
\begin{center}
\includegraphics[width=10cm]{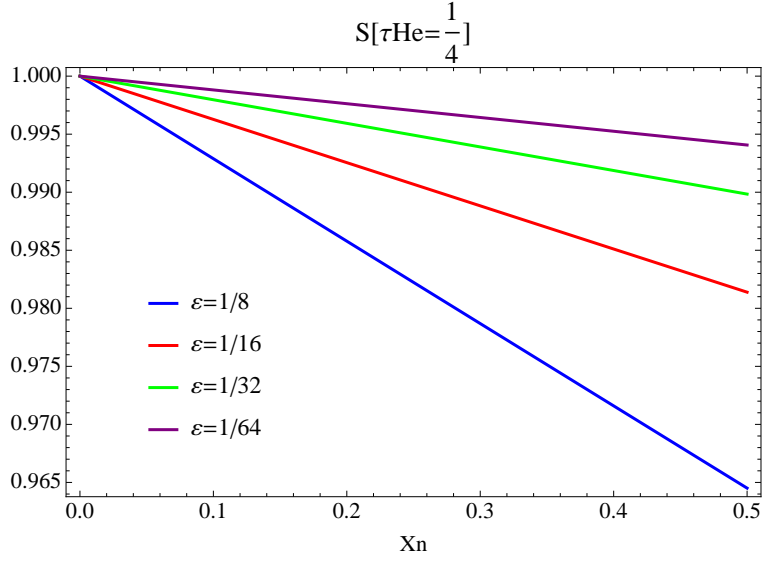}
\caption{The radiation produced by the baryons as a function of $X_{n}$. We can see that $S_{rad}$ depends linearly on $X_{n}$ and that its amount decreases if $\epsilon$ becomes large.}
\label{lin}
\end{center}
\end{figure}
Although we have given the differential equations to calculate $S_{rad}$, the $v_{h}$ dependence is almost the same as that of the simplified case(\ref{simplerad}). The problem is how  atomic nuclei change it. To see this, we first need to know the $X_{n}$ dependence of the radiation, and this can be checked by solving the differential equations numerically. See Fig.\ref{lin}. One can see that the radiation depends on $X_{n}$ almost linearly. Thus, we can write $S_{rad}$ as
\be S_{rad}= constant \times\left(\frac{1}{M_{pl}^{2}}\right)^{\frac{1}{3}}\times (N_{B}m_{p})^{\frac{4}{3}}\tau^{\frac{2}{3}}_{p} \times\left(1-c(\epsilon,\frac{\tau_{He}}{\tau_{p}},\frac{m_{He}}{m_{p}})X_{n}\right),\label{apprad2}\e
where the coefficient $c$ is a function of $\epsilon,\frac{\tau_{He}}{\tau_{p}}$ and $\frac{m_{He}}{m_{p}}$ because we can rewrite the differential equations in terms of them if we rescale $a$ and $t$ as
\be a(t)\rightarrow a^{'}(t):=\frac{a(t)}{m_{p}^{\frac{1}{3}}}\h{2.5mm},\h{2.5mm}t\rightarrow t^{'}:=\frac{t}{\tau_{p}}.\e
$c$ represents how much the radiation is reduced by the existence of neutrons. The important fact is that, since the five parameters $m_{p},m_{He},\tau_{p},\tau_{He},\epsilon$ should be determined by the nuclear physics, we can treat $c$ as a constant when we fix $\Lambda_{QCD}$ and the current quark masses. This is the reason why we fix the current quark masses in the next subsection. In this case, the $v_{h}$ dependence of $S_{rad}$ is given by
\be S_{rad}= constant\times\underbrace{N_{B}^{\frac{4}{3}}\times(1-c\cdot X_{n}).}_{\text{$v_{h}$ dependence when $m_{q}=$fixed}}\label{apprad3}\e
In the previous section, we have studied the behavior of $X_{n}$ and $N_{B}$ as a function of $v_{h}$. Thus, one can easily understand why Eq.(\ref{apprad3}) has a maximum around $v_{h}=246$GeV as follows. In the region of small $v_{h}<200$GeV, since $X_{n}$ is approximately zero as is shown in Fig.\ref{xngraph}, we can write
\be S_{rad}\propto N_{B}^{\frac{4}{3}},\e
where $N_{B}$ is an increasing function of $v_{h}$ as is shown in Fig.\ref{baryon1}. Thus, $S_{rad}$ is a increasing function of $v_{h}$ in this region. On the other hands, in the region of $v_{h}>>200$GeV, since $X_{n}$ increases and $N_{B}$ is almost constant, the radiation decreases due to the effect of $c\cdot X_{n}$. 

\subsection{The Numerical Result}\label{subsec:numerical}
\begin{figure}
\begin{center}
\begin{tabular}{c}
\begin{minipage}{0.5\hsize}
\begin{center}
\includegraphics[width=10cm]{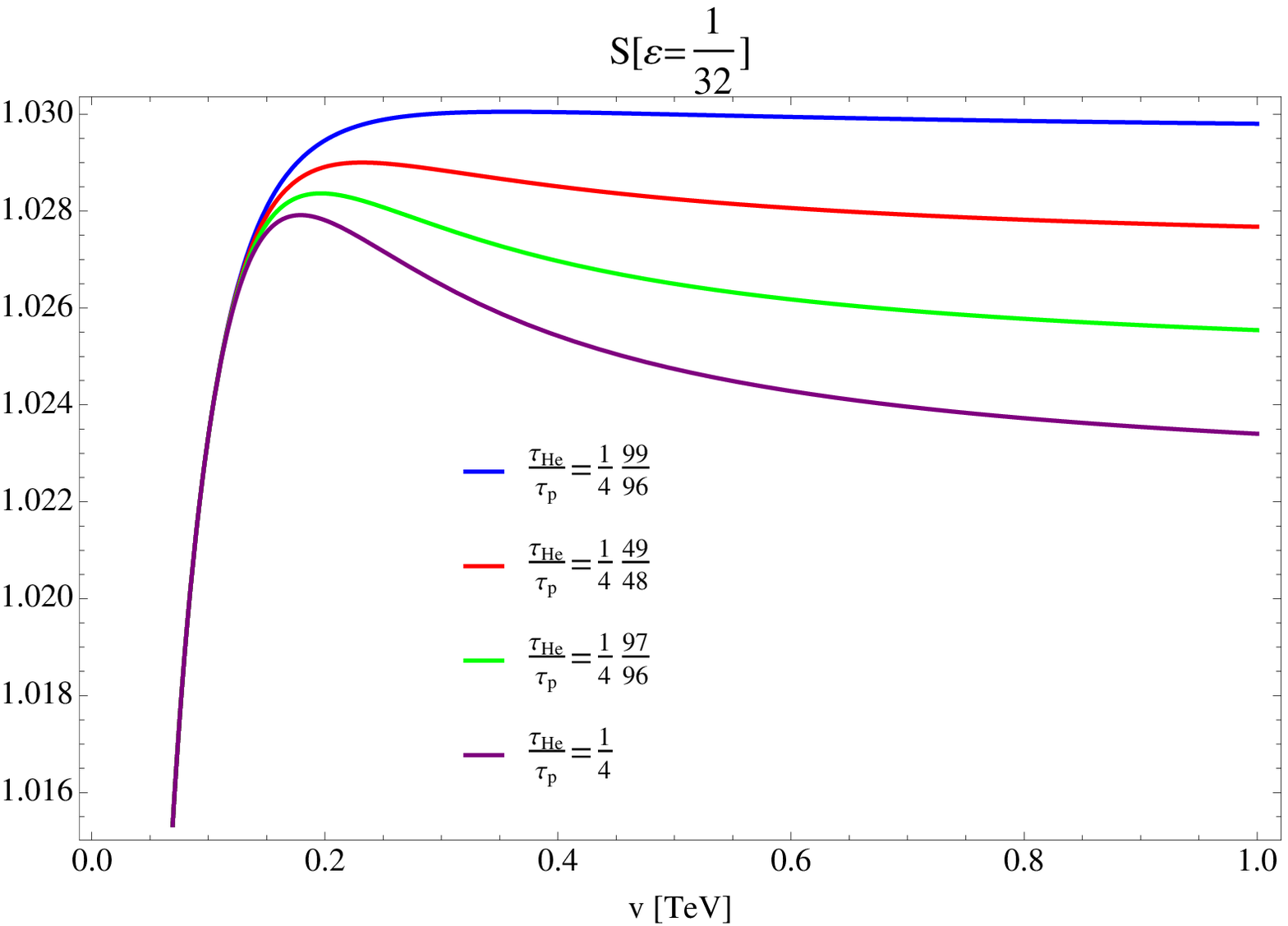}
\end{center}
\end{minipage}
\\
\\
\begin{minipage}{0.5\hsize}
\begin{center}
\includegraphics[width=10cm]{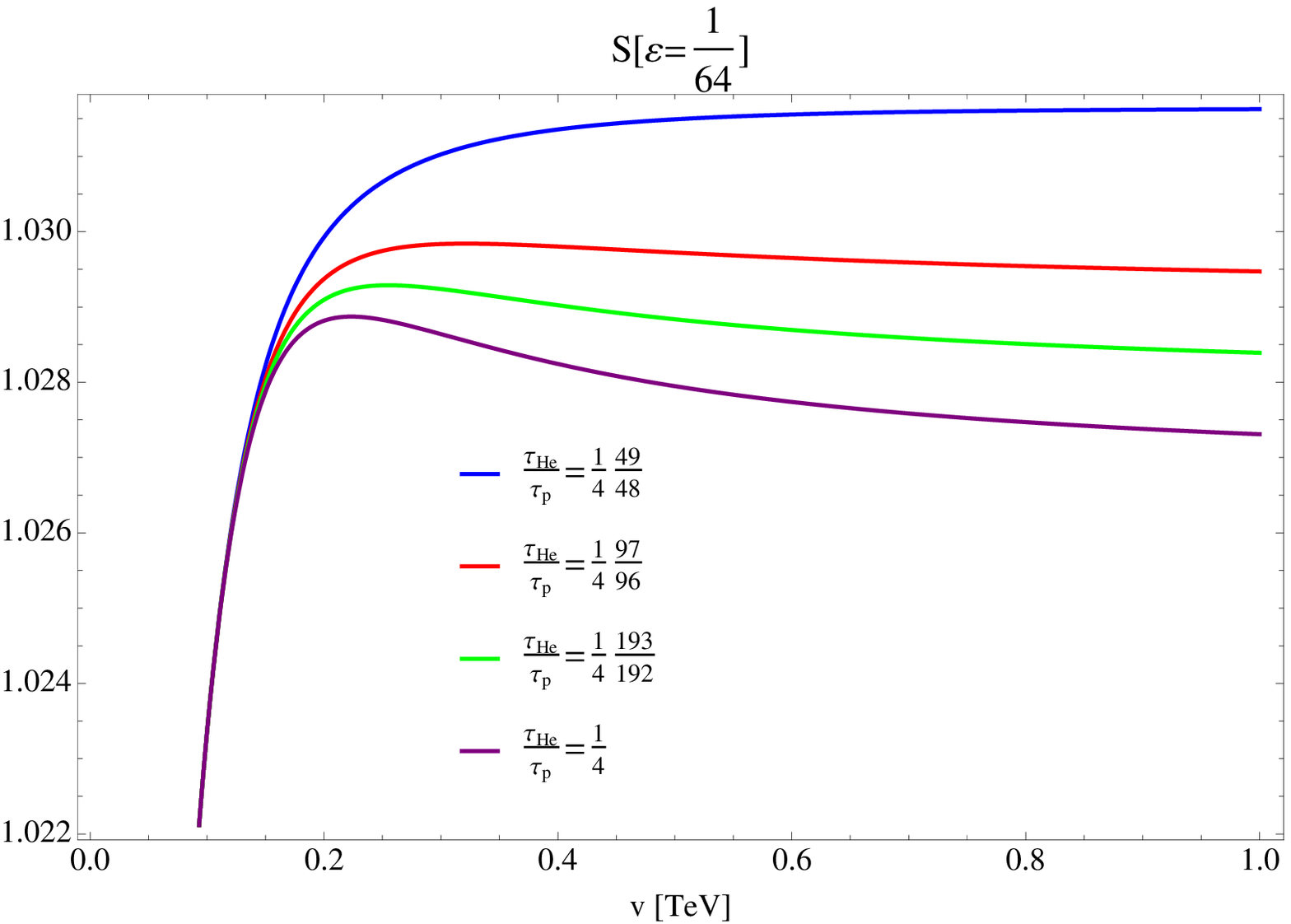}
\end{center}
\end{minipage}
\end{tabular}
\caption{The radiation as a function of $v_{h}$ when the gauge couplings, the Higgs self coupling and the current quark masses are fixed. The upper(lower) graph shows the case  $\epsilon=\frac{1}{32}(\frac{1}{64})$ for various values of $\frac{\tau_{He}}{\tau_{p}}$. We can see that $S_{rad}$ has a global maximum around $v_{h}=246$GeV without the fine tunings of $\tau_{He}$ and $\epsilon$.}
\label{entmass}
\end{center}
\end{figure}
As discussed in the previous subsection, the reason why we fix the current quark masses is that we need not to consider  complicated effects of nuclear physics. Fig.\ref{entmass} shows the numerical results at a sufficiently late time ($t=20\tau_{p}$) by using the differential equations in Section \ref{sec:baryon history}. One can see that $S_{rad}$ has a global maximum around $v_{h}\simeq200$GeV if the life time of a helium nucleus $\tau_{He}$ is not too big. Since $\tau_{He}$ is expected to be within (see \ref{app:epsilon})
\be \frac{1}{4}\tau_{p}<\tau_{He}<\frac{1}{4}\times\frac{7}{6}\tau_{p},\e
we can conclude that the radiation of our universe has a global maximum very generally.

\begin{figure}[!h]
\begin{center}
\includegraphics[width=9cm]{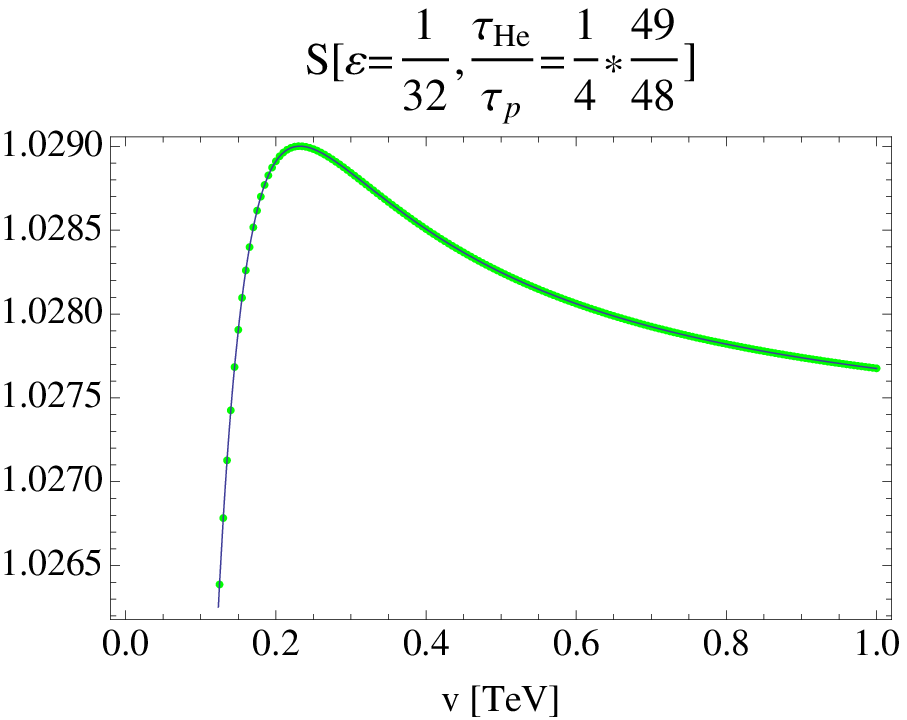}
\caption{The fitting result when $\epsilon=\frac{1}{32},\frac{\tau_{He}}{\tau_{p}}=\frac{1}{4}\times\frac{49}{48}$ and $\frac{m_{He}}{m_{p}}=4-\frac{28}{938}$. The green line shows the result of the differential equations, and the black line shows the fit by Eq.(\ref{eq:fit}) with the best fit value $c=0.010$. They overlap almost completely.}
\label{fit}
\end{center}
\end{figure}
\noindent We can also see how precisely the approximate formula
\be S_{rad}\propto N_{B}^{\frac{4}{3}}\times(1-c\cdot X_{n})\label{eq:fit}\e 
reproduces the above numerical result. For example, Fig.\ref{fit} shows the case of the following parameters:
\be \epsilon=\frac{1}{32}\h{2.5mm},\h{2.5mm}\frac{\tau_{He}}{\tau_{p}}=\frac{1}{4}\times\frac{49}{48}\h{2.5mm},\h{2.5mm}\frac{m_{He}}{m_{p}}=4-\frac{28}{938}.\e
One can see that the numerical result (the green line) of the differential equations is almost completely reproduced by Eq.(\ref{eq:fit}) if we choose the best fit value $c=0.010$ (the black line). Note that the coefficient $c$ is around
\be c\simeq {\cal{O}}(0.01).\e
The problem is how such a small value is obtained. By solving the differential equations for various initial conditions and parameters, we can read the behavior of $c$:
\be c\left(\epsilon,\frac{\tau_{He}}{\tau_{p}},\frac{m_{He}}{m_{p}}\right)=0.545\times\epsilon-1.829\times\left(\frac{\tau_{He}}{\tau_{p}}-\frac{1}{4}\right)+0.118\times\left(4-\frac{m_{He}}{m_{p}}\right)+\cdots.\label{funcc}\e
One can see that while $\epsilon$ and the binding energy $4-\frac{m_{He}}{m_{p}}$ have effects to decease the radiation, $\frac{\tau_{He}}{\tau_{p}}-\frac{1}{4}$ has an increasing effect. Thus, it is not so unnatural for $c$ to become ${\cal{O}}(0.01)$. Actually, we can check that $c$ stays of order $0.01$ when we change $\epsilon,\frac{\tau_{He}}{\tau_{p}}$ and $\frac{m_{He}}{m_{p}}$ within the physically reasonable region. Therefore, we can conclude that the QCD naturally gives the above small value, and the maximum entropy principle is justified for $v_{h}$.

\section{Summary and Discussion}
\begin{figure}[!h]
\begin{center}
\includegraphics[width=9cm]{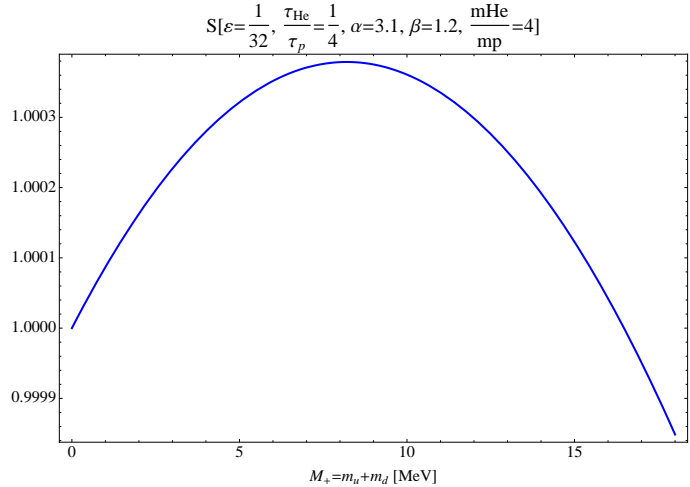}
\caption{The radiation of the universe as a function of the sum of $m_{u}$ and $m_{d}$. Here, we have used the approximate formulas which are given by Eq.(\ref{eq:appformula1})(\ref{eq:proton life}), and fixed $v_{h}$ and $y_{d}-y_{u}$ at the observed values.}
\label{approximate radiation}
\end{center}
\end{figure}
In this paper, we have shown that the radiation at the late stage of the universe becomes maximum around $v_{h}=246$GeV assuming that the baryon number is produced by the sphaleron process and that the current quark masses $m_{i}$, the gauge copings and the Higgs self coupling are fixed. Since the radiation of the universe can be regarded as the entropy, this conclusion is the maximum entropy principle. The reason why we have fixed $m_{i}$ is that we can examine the $v_{h}$ dependence without going through the complicated nuclear physics. Actually the maximum of $S_{rad}$ is determined by the behavior of $N_{B}$ and $X_{n}$ as functions of $v_{h}$. We have seen that the existence of atomic nuclei plays a very important role to maximize $S_{rad}$. In other words, the maximum entropy principle forces the existence of atomic nuclei, and as a consequence, the Higgs expectation value is fixed to around $246$GeV. This is reminiscent of the anthropic principle \cite{Martel:1997vi,Agrawal:1997gf,Agrawal:1998xa} in which the existence of human beings requires atomic nuclei. In our case, however, they are naturally required by the fundamental low.  
\\

Although what we actually want to do is to regard $S_{rad}$ as a function of the SM parameters, and confirm the maximum entropy principle by changing them independently, in this paper, we have considered only $v_{h}$ so that we need not to consider the effects of nuclear physics. However, by using the approximate formulas such as
\be m_{p}=\alpha\Lambda_{QCD}+\beta(2m_{u}+m_{d})\h{2.5mm},\h{2.5mm}m_{\pi}^{2}=\gamma\Lambda_{QCD}\frac{m_{u}+m_{d}}{2},\label{eq:appformula1}\e
\be  \Gamma_{p}=\tau_{p}^{-1}\propto g^{2}\frac{m_{p}^{5}}{M_{X}^{4}}\times\left(1-\frac{m_{\pi}^{2}}{m_{p}^{2}}\right)^{2},\label{eq:proton life}\e
it may be possible to determine also $y_{u}$ and $y_{d}$. For example, Fig.\ref{approximate radiation} shows the numerical result of $S_{rad}$ as a function of $m_{+}:=m_{u}+m_{d}=(y_{u}+y_{d})\cdot v_{h}/\sqrt{2}$ when the difference of the Yukawa couplings $y_{-}=y_{d}-y_{u}$ and $v_{h}$ are fixed at the observed values. Here, we have used  the ad-hoc values
\be \Lambda_{QCD}=300\text{MeV}\h{2mm},\h{2mm}\alpha=3.1\h{2mm},\h{2mm}\beta=1.2\h{2mm},\h{2mm}\frac{\tau_{He}}{\tau_{p}}=\frac{1}{4}\h{2mm},\h{2mm}\frac{m_{He}}{m_{p}}=4\h{2mm},\h{2mm}\epsilon=\frac{1}{32},\e
and experimental value\footnote{This can be used to determine $\gamma\times\Lambda_{QCD}$.}
\be m_{\pi}|_{\text{phys}}=135\text{MeV}.\e
One can see that $S_{rad}$ has a maximum around the experimental values. Furthermore, since $S_{rad}$ is an increasing function of $y_{-}$ if we use the above equations(see the qualitative equation (\ref{apprad3})), we may conclude that the Yukawa couplings are determined by the maximum entropy principle at 
\be(y_{u},y_{d})\simeq(0, 4\times10^{-5}).\e
Making the above argument more precisely is the future work. Although the understanding of the maximum entropy principle is in a primitive level at present, it is very interesting to study whether the radiation of the universe at the late stage is really decreased  when we change each of the SM parameters independently. 

\section*{Acknowledgement} 
We thank Takahiro Kubota and Hiroshi Itoyama for valuable discussions. The work of Y.~H. is supported by a Grant-in-Aid for Japan Society for the Promotion of  Science(JSPS) Fellows No.25$\cdot$1107.

\appendix 
\def\thesection{Appendix \Alph{section}}
\section{ WKB Solution  }\label{WKB}
\def\thesection{ \Alph{section}}

Eigenstates of the wave function of the universe are given by
\be \hat{{\cal{H}}}\cdot\phi_{E}(a)=\left(-\frac{\hat{p}^{2}_{a}}{2M_{pl}^{2}a}+\frac{a^{3}\rho(a)}{6}\right)\cdot\phi_{E}(a)=E\phi_{E}(a)\e
\be \leftrightarrow \left(-\frac{1}{2}\hat{p}_{a}^{2}-V(a)\right)\phi_{E}(a)=M_{pl}^{2}aE\phi_{E}(a)\e
\be V(a):=-\frac{M_{pl}^{2}a^{4}\rho(a)}{3}=-\frac{M_{pl}^{2}}{3}\times a^{4}\times\left(\frac{M}{a^{3}}+\frac{R}{a^{4}}-\frac{M_{pl}^{2}}{a^{2}}+M_{pl}^{2}\Lambda\right)\e
where $p_{a}=-M_{pl}^{2}a\dot{a}$. According to the ordinary quantum mechanics, the WKB wave function is given by
\be \phi_{E}(a)\propto\exp\left(\frac{i}{\hbar}S_{0}+iS_{1}+o(\hbar)\right).\label{WKBfunction0}\e
We have to pay attention to the order of $a$ and $\hat{p}_{a}$ because the Hamiltonian has the kinetic term that depends on the scale factor $a$. The correct order is given by 
\be \frac{p^{2}_{a}}{a}=-\hbar^{2}\frac{d}{da}a\frac{d}{da}=-\hbar^{2}\left(\frac{1}{a}\frac{d^{2}}{da^{2}}-\frac{1}{a^{2}}\frac{d}{da}\right).\label{order}\e
If we assume this order, the WKB wave function (\ref{WKBfunction0}) satisfies the following Schr$\ddot{\text{o}}$dinger equation:
\be -\frac{1}{2}\left(\frac{d(S_{0}+\hbar S_{1})}{da}\right)^{2}+\frac{i\hbar}{2}\frac{d^{2}(S_{0}+\hbar S_{1})}{da^{2}}-\uwave{\frac{i\hbar}{2a}\frac{dS_{0}}{da}}=M_{pl}^{2}aE+V(a),\e
which leads to
\be {\cal{O}}(\hbar^{0}):-\left(\frac{dS_{0}}{da}\right)^{2}=2\left(M_{pl}^{2}aE+V(a)\right):=-p^{2}_{cl},\e
\be {\cal{O}}(\hbar):-2\frac{dS_{0}}{da}\cdot\frac{dS_{1}}{da}+i\left(\frac{d^{2}S_{0}}{da^{2}}-\uwave{\frac{1}{a}\frac{dS_{0}}{da}}\right)=0.\e
Solutions to this equations are
\be S_{0}=\int^{a} da^{'}p_{cl}\h{3mm},\h{3mm}S_{1}=\frac{i}{2}\left(\log p_{cl}-\log a\right).\e
Then, except for the normalization factor, Eq.(\ref{WKBfunction0}) becomes 
\be \phi_{E}\propto\exp\left(\frac{i}{\hbar}S_{0}+iS_{1}+o(\hbar)\right)=\frac{\uwave{\sqrt{a}}}{\sqrt{p_{cl}}}\exp\left(\frac{i}{\hbar}\int^{a} da^{'}p_{cl}\right).\label{WKBfunction1}\e
Note that the factor $\sqrt{a}$ is different from the ordinary WKB solution. For large $a$, the classical momentum becomes
\be p_{cl}=\sqrt{-2\left(V(a)+M_{pl}^{2}aE\right)}\simeq\sqrt{-2M_{pl}^{2}\left(-\frac{M_{pl}^{2}\Lambda a^{4}}{3}+aE\right)}\simeq\sqrt{\frac{2\Lambda}{3}}\cdot (aM_{pl})^{2}-\sqrt{\frac{3}{2\Lambda}}\cdot\frac{E}{a},\e
and the corresponding WKB solution (\ref{WKBfunction1}) is
\be \phi_{E}\simeq\left(\frac{3}{2\Lambda a^{2}}\right)^{\frac{1}{4}}\times\exp\left[\frac{i}{\hbar}\left(\sqrt{\frac{2\Lambda}{3}}\frac{M_{pl}^{2} a^{3}}{3}-\sqrt{\frac{3}{2\Lambda}}E\log a\right)\right].\e
Finally we can obtain the correct inner product between these eigenstates:
\begin{align} (\phi_{E},\phi_{E^{'}})&\propto\sqrt{\frac{3}{2\Lambda}}\times\int da\frac{\exp\left[\frac{i}{\hbar}\sqrt{\frac{3}{2\Lambda}}(E-E^{'})\log a\right]}{a}\nonumber\\
&=\hbar\delta\left(E-E^{'}\right).\label{inner1}\end{align}
From Eq.(\ref{inner1}), we can see the normalization factor as $\hbar^{\frac{1}{2}}$. Thus, the correct WKB solution is
\be \phi_{E}=\sqrt{\frac{a}{\hbar p_{cl}}}\times\exp\left(\frac{i}{\hbar}\int^{a} da^{'}p_{cl}\right).\label{WKB solution}\e

\def\thesection{Appendix \Alph{section}}
\section{Probabilistic Interpretation of the Wave Function of the Multiverse}\label{Multiverse}
In this Appendix, we review the previous works \cite{Kawai:2011rj,Kawai:2011qb} that interpret the squared absolute value of the wave function of the multiverse as the probability distribution. See also \cite{Klebanov:1988eh,Polchinski:1988ua,Giddings:1988wv,Nielsen:1988kf,Grinstein:1988wr,Preskill:1989zu,Rubakov:1988jf,Strominger:1988yt,Fischler:1989ka,Cline:1989vj}. 
\subsection{Multiverse Path Integral}
Coleman's first idea is that, by taking the wormhole configurations into account, we can obtain the multiverse path integral. We start with the Euclidean Einstein Gravity,
\be Z=\sum_{all{\cal{M}}}\int_{{\cal{M}}} {\cal{D}}g{\cal{D}}\phi \hspace{2mm}\exp(-S^{G}_{E}-S^{M}_{E})\e
\be S^{G}_{E}=-\frac{1}{16\pi G}\int d^{4}x\sqrt{g(x)}(R+2\Lambda_{o})\e
\be S_{E}^{M}=\int d^{4}x\sqrt{g(x)}{\cal{L}}(\phi,\partial\phi),\e
where ${\cal{M}}$ represents a four dimensional manifold, $\Lambda_{o}$ is the bare cosmological constant and $\phi$ is matter field. Classical solution of the Euclidean Gravity is called the wormhole and Coleman discussed the dilute gas approximation of the wormhole configurations, and obtained
\begin{align}Z=\sum_{n=0}^{\infty}\int_{{\cal{M}}_{n}}{\cal{D}}g{\cal{D}}\phi\exp\{\sum_{i,j}c_{ij}\int dx^{4}dy^{4}&\sqrt{g(x)}\sqrt{g(y)}{\cal{O}}(x)_{i}{\cal{O}}(y)_{j}\exp(-2S_{W})\}\nonumber\\
&\times\exp(-S_{E}^{G}-S_{E}^{M}).\end{align}
where $S_{W}$ is the classical action of a wormhole and ${\cal{M}}_{n}$ represents a manifold such that there are n universes, which are disconnected each other. In fact, this result takes into account only the wormhole configurations having two legs. Including all the different types of wormholes, we obtain the following multi local effective action\footnote{We can also obtain the multi local action in matrix model. See \cite{Asano:2012mn}.}:
\be Z=\sum_{n=0}^{\infty}\int_{{\cal{M}}_{n}} {\cal{D}}g{\cal{D}}\phi \exp(-S^{G}_{E}-S^{M}_{E}-\sum_{k=2}^{\infty}Z_{k})
\label{multipath1}\e
where $Z_{k}$ comes from wormholes having k legs;
\be Z_{k}:=\sum_{i_{1},i_{2}\cdots,i_{k}}c_{i_{1}i_{2}\cdots i_{k}}\prod_{j=1}^{k}\int d^{4}x_{j}\sqrt{g(x_{j})}{\cal{O}}^{i_{j}}(x_{j})
\exp(-S_{W}):=\sum_{i_{1},i_{2}\cdots,i_{k}}c^{'}_{i_{1}i_{2}\cdots i_{k}}S_{i_{1}}S_{i_{2}}\cdots S_{i_{k}}.\label{multilocal}\e
Here, 
\be S_{i}:=\int dx^{4}\sqrt{g(x)}{\cal{O}}^{i}(x)\e
is the ordinary local action. We use the Lorentzian counterpart of Eq.(\ref{multilocal}) by the Wick rotation:
\be\sum_{n=0}^{\infty}\int_{{\cal{M}}_{n}} {\cal{D}}g{\cal{D}}\phi \exp(iS^{G}_{E}+iS^{M}_{E}+i\sum_{k=2}^{\infty}Z_{k}):=\sum_{n=0}^{\infty}\int_{{\cal{M}}_{n}} {\cal{D}}g{\cal{D}}\phi \exp(iS_{eff})\label{multipath2}\e
\be S_{eff}=S^{G}_{E}+S^{M}_{E}+\sum_{ij}c_{ij}S_{i}S_{j}+\sum_{ijk}c_{ijk}S_{i}S_{j}S_{k}+\cdots.\label{effective action}\e
By making the Fourier transform, we can write Eq.(\ref{multipath2}) as the usual path integral form:
\begin{align} Z&=\sum_{K}\int d\overrightarrow{\lambda}\omega(K,\lambda_{1},\lambda_{2},\cdots)\sum_{n=0}^{\infty}\int_{{\cal{M}}_{n}}{\cal{D}}g{\cal{D}}\phi \hspace{1mm}\exp(i\sum_{i}\lambda_{i}S_{i})\nonumber\\
&:=\sum_{K}\int d\overrightarrow{\lambda}\omega(K,\lambda_{1},\lambda_{2},\cdots)\sum_{n=0}^{\infty}\frac{1}{n!}(Z_{universe}^{(\lambda)})^{n}.\label{multipath3}\end{align}
Here,
\be Z_{universe}^{(\lambda)}:=\int{\cal{D}}g{\cal{D}}\phi \hspace{1mm}\exp(i\sum_{i}\lambda_{i}S_{i})\label{unipath1}\e
is the ordinary path integral of the universe, and $\omega(K,\{\lambda_{i}\})$ is the Fourier coefficient and K represents the topology of a single universe. One can see that $\{\lambda_{i}\}$ serve as coupling constants of a single universe. Eq.(\ref{multipath3}) indicates that they are variables ( not the constants ). 
In the following discussion, we assume the two conditions besides the Wick rotation:\\
\begin{enumerate}\item We consider the homogeneous, isotropic universe with $S^{3} (K=1)$ topology:
\begin{equation} d^{2}s=-N(t)d^{2}t+a^{2}(t)\left(d\mathbf{x}^{2}+\frac{(\mathbf{x}\cdot d\mathbf{x})^{2}}{1-\mathbf{x}^{2}}\right).\end{equation}
\item Matter and radiation fields are included as the potential of the scale factor $a(t)$, namely the Hamiltonian is given by
\be {\cal{H}}(\lambda)=-\frac{1}{2M_{pl}^{2}a}p^{2}_{a}+a^{3}\rho(a).\e
where $\rho(a)$ is the energy density of a single universe. 
\end{enumerate}
Based on this assumptions, Eq.(\ref{unipath1}) becomes
\be Z^{(\lambda)}_{universe}(a_{f},a_{i})=\int{\cal{D}}p_{a}\int_{t=0 ,a(0)=a_{i}}^{t=1 ,a(1)=a_{f}}{\cal{D}}a{\cal{D}}N \exp\{i\int_{0}^{1} dt(p_{a}\dot{a}-N{\cal{H}}(\lambda))\}. \label{unipath2}\e
where we have written the boundary condition explicitly. By choosing the gauge such that $N(t)=T$(constant) and rescaling t as $t\rightarrow t\times T$, Eq.(\ref{unipath2}) becomes
\begin{align} Z^{(\lambda)}_{universe}(a_{f},a_{i})&=\int_{-\infty}^{\infty} dT\int_{t=0 ,a(0)=a_{i}}^{t=T ,a(T)=a_{f}}{\cal{D}}a \exp\{i\int_{0}^{T} dt(p_{a}\dot{a}-{\cal{H}}(\lambda))\}\nonumber\\
&=\int_{\infty}^{\infty}dT\langle a_{f}|\exp(-i{\cal{\hat{H}}}(\lambda))|a_{i}\rangle\nonumber\\
&=\langle a_{f}|\delta({\cal{\hat{H}}}(\lambda))|a_{i}\rangle. \label{unipath3}\end{align}
If the cosmological constant $\Lambda(\{\lambda_{i}\})$ is positive, the eigenvalues of ${\cal{H}}(\lambda)$ are continuous, so we can choose the complete set as follows:
\be \langle \phi_{E}|\phi_{E^{'}}\rangle=\delta(E-E^{'}),\e
\be {\cal{\hat{H}}}(\lambda)|\phi_{E}\rangle=E|\phi_{E=0}\rangle,\e
\be 1=\int_{-\infty}^{\infty}|E\rangle\langle E|.\label{complete}\e
Inserting Eq.(\ref{complete}) to Eq.(\ref{unipath3}), we obtain
\begin{align} &=\langle a_{f}|\delta({\cal{\hat{H}}}(\lambda))\int_{-\infty}^{\infty}|E\rangle\langle E||a_{i}\rangle\nonumber\\
&=\phi_{E=0}(a_{f})\phi^{*}_{E=0}(a_{i}).\label{unipath4}\end{align}
On the other hands, when $\Lambda(\{\lambda_{i}\})$ is negative, the complete set becomes
\be \langle \phi_{E}|\phi_{E^{'}}\rangle=\delta_{E,E^{'}},\e
\be {\cal{\hat{H}}}(\lambda)|\phi_{E}\rangle=E|\phi_{E=0}\rangle,\e
\be 1=\sum_{E}|E\rangle\langle E|,\label{complete2}.\e
because the eigenvalues of ${\cal{\hat{H}}}(\lambda)$ are discrete. As well as the $\Lambda(\{\lambda_{i}\})>0$ case, the path integral given by Eq.(\ref{unipath3}) becomes the same expression as Eq.(\ref{unipath4}). 

\subsection{Multiverse Wave Function and Probabilistic Interpretation}
In this subsection, we construct the multiverse wave function. Because we have to treat the coupling constants $\{\lambda_{i}\}$ of the universe as variables, the quantum state of the n universes is given by
\be |\Psi_{n},\{\lambda_{i}\}\rangle=\frac{1}{\sqrt{n!}}\mu_{K}^{n}|\phi_{universe}\rangle\otimes\cdots\otimes|\phi_{universe}\rangle\otimes\omega(\lambda)|\{\lambda_{i}\}\rangle,\e
where $|\phi_{universe}\rangle$ is the state of a single universe, and $\mu_{K}$ is the probability amplitude of a universe emerging from nothing. Here we have divided $|\Psi_{n},\{\lambda_{i}\}\rangle$ by the factor $\sqrt{n!}$ because we can not distinguish each universe.  We can now express the wave function of the n universes as
\be \Psi(a_{1},a_{2},\cdots,a_{n},\{\lambda_{i}\})=\frac{1}{\sqrt{n!}}\mu_{K}^{n}\cdot\omega(\lambda)\times\phi_{universe}(a_{1},\{\lambda_{i}\})\cdot\phi_{universe}(a_{2},\{\lambda_{i}\})\cdots\phi_{universe}(a_{n},\{\lambda_{i}\})\label{eq:multiwave}\e
where 
\be \phi_{universe}(a,\{\lambda_{i}\}):=\langle a|\phi_{universe}\rangle= \int da^{'}\h{1mm}Z^{(\lambda)}_{universe}(a,a^{'})\langle a^{'}|\phi_{universe}\rangle\label{uniwave1}\e
is the wave function of a single universe. We interpret Eq.(\ref{eq:multiwave}) as follows:\\
\\
\underline{{\it{\textbf{Probabilistic Interpretation of Wave Function of Multiverse}}}}\\
We interpret that 
\be |\Psi(a_{1},a_{2},\cdots,a_{n},\{\lambda_{i}\})|^{2}\e
represents the probability that  n universes having coupling constants $\{\lambda_{i}\}$ and scale factors $(a_{1},a_{2},\cdots,a_{n})$ are observed.
\\
\\
We can now obtain the probability distribution $P(\{\lambda_{i}\})$ by tracing out the number of universes and the scale factors $a_{i}$ :
\begin{align} P(\{\lambda_{i}\})&=\sum_{n=0}^{\infty}\int\cdots\int \prod_{k=1}^{n}da_{k}|\Psi(a_{1},a_{2},\cdots,a_{n},\{\lambda_{i}\})|^{2}\nonumber\\
&=\sum_{n=0}^{\infty}\frac{1}{n!}|\mu_{K}|^{2n}\cdot|\omega(\lambda)|^{2}\cdot\prod_{k=1}^{n}\left(\int da_{k}|\phi_{universe}(a_{k},\{\lambda_{i}\})|^{2}\right)\nonumber\\
&=|\omega(\lambda)|^{2}\times\exp\left(|\mu_{K}|^{2}\cdot\int da|\phi_{universe}(a,\{\lambda_{i}\})|^{2}\right)\label{pro1}.\end{align}
To analyze Eq.(\ref{pro1}), we must determine the wave function of the single universe $\phi_{universe}(a,\{\lambda_{i}\})$. In this paper, we assume that a universe emerges from nothing with a small size $\epsilon$:
\be |\phi_{universe}\rangle=|\epsilon\rangle.\label{initial}\e
Generally speaking, there is no restriction  on $|\phi_{universe}\rangle$. For the present, Eq.(\ref{initial}) is our proposal for the quantum state of the universe. 

By this assumption, the wave function of the universe (\ref{uniwave1}) becomes
\begin{align} \phi_{universe}(a,\{\lambda_{i}\})&=\int da^{'}\h{1mm}Z^{(\lambda)}_{universe}(a,a~{'})\langle a^{'}|\epsilon\rangle\nonumber\\
&=Z^{(\lambda)}_{universe}(a,\epsilon)=\phi_{E=0}(a)\phi^{*}_{E=0}(\epsilon),\end{align}
where we have used Eq.(\ref{unipath4}). Here, note that $\phi_{E=0}(a)$ depends on $\{\lambda_{i}\}$. 

Now we obtain the explicit expression of $P(\{\lambda_{i}\})$:
\be P(\{\lambda_{i}\})=|\omega(\lambda)|^{2}\times\exp\left(|\mu_{K}\phi_{E=0}(\epsilon)|^{2}\cdot\int da|\phi_{E=0}(a)|^{2}\right).\label{pro2}\e
The intuitive understanding of this equation is as follows. By using Eqs.(\ref{unipath3})(\ref{unipath4}) and regularizing the time integral by the cut off $T_{Max}$, the integral included in Eq.(\ref{pro2}) becomes
\begin{align} |\phi_{E=0}(\epsilon)|^{2}\int da|\phi_{E=0}(a)|^{2}&=\int da\h{1mm}Z^{(\lambda)*}_{universe}(a,\epsilon)\times Z^{(\lambda)}_{universe}(a,\epsilon)\nonumber\\
&=\int da \langle\epsilon|\int_{-T_{Max}}^{T_{Max}}dTe^{-i{\cal{H}}(\lambda)}|a\rangle\langle a|E=0\rangle\times\langle \epsilon|E=0\rangle\nonumber\\
&=\int_{-T_{Max}}^{T_{Max}}dT\langle\epsilon|E=0\rangle\langle\epsilon|E=0\rangle\nonumber\\
&=2T_{Max}\times|\phi_{E=0}(\epsilon)|^{2}.\label{lifetimeuniverse}\end{align}
Thus, Eq.(\ref{pro2}) has a strong peak where the life time of the universe becomes maximum. In conclusion, the dynamics of the multiverse and wormholes naturally fixes the low energy parameters $\{\lambda_{i}\}$ in such a way that the life time of the universe becomes maximum.

\subsection{The Maximum Entropy Principle}
\begin{figure}
\begin{center}
\includegraphics[width=9cm]{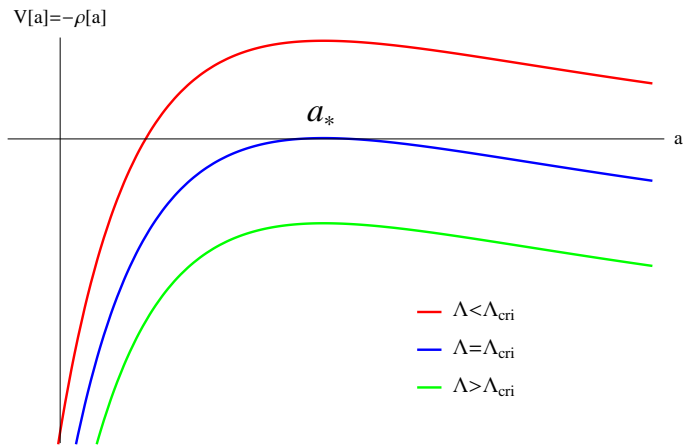}
\caption{The potentials of the $S^{3}$ universe for the various values of the Cosmological Constant $\Lambda$. If $\Lambda=\Lambda_{cr}$, the universe spends infinite times to grow up to the size $a_{*}$.}
\label{pot1}
\end{center}
\end{figure}
In this section, we show how the maximum entropy principle is deduced for the $S^{3}$ universe by using Eq.(\ref{pro2}). Eq.(\ref{pro2}) has the following $\{\lambda_{i}\}$ dependences:
\be \omega(\lambda)\h{2mm},\h{2mm}|\phi_{E=0}(\epsilon)|^{2}\h{2mm},\h{2mm}\int da|\phi_{E=0}|^{2}.\e
Among them, $\omega(\lambda)$ and $|\phi_{E=0}(\epsilon)|^{2}$ are not so important because they do not have a sharp peak. On the other hands, as we will see, the integral
\be \int da|\phi_{E=0}(a)|^{2}\label{uniwavefunc}\e
depends on $\{\lambda_{i}\}$ most strongly. The hamiltonian of the $S^{3}$ universe is given by (see \ref{WKB})
\be {\cal{H}}(\lambda)=-\frac{1}{2}p_{a}^{2}-V(a),\e
where
\be V(a):=-\frac{M_{pl}^{2}a^{4}\rho(a)}{3}=-\frac{M_{pl}^{2}}{3}\times a^{4}\times\left(\frac{M}{a^{3}}+\frac{R}{a^{4}}-\frac{M_{pl}^{2}}{a^{2}}+M_{pl}^{2}\Lambda\right)\e
is the potential of the universe. As discussed in \cite{Kawai:2011rj}\cite{Kawai:2011qb}, there is the critical surface of codimension one on which we have
\be \Lambda_{cr}(\{\lambda_{i}\}_{cr})\simeq\begin{cases}\frac{M_{pl}^{2}}{R(\lambda_{cr})}&\text{(for radiation dominated universe)}\\
\\ \frac{M_{pl}^{4}}{M(\lambda_{cr})^{2}}&\text{(for matter dominated universe)}\end{cases}\e
such that the maximum value of the potential $V(a^{*})$ becomes 0 (see Fig.\ref{pot1}). In this case, the integral (\ref{uniwavefunc}) becomes 
\be \int da|\phi_{E=0}(a)|^{2}\simeq\begin{cases}\frac{R^{\frac{1}{2}}(\lambda_{cr})}{M_{pl}}\log(\frac{R^{\frac{1}{2}}(\lambda_{cr})}{M_{pl}\Delta a_{WKB}})+\frac{R^{\frac{1}{2}}(\lambda_{cr})}{M_{pl}}\log(\frac{a_{IR}}{a^{*}})&\text{(for radiation dominated universe)}\\
\\ \frac{M(\lambda_{cr})}{M^{2}_{pl}}\log(\frac{M(\lambda_{cr})}{\Delta a_{WKB}})+\frac{M(\lambda_{cr})}{M^{2}_{pl}}\log(\frac{a_{IR}}{a^{*}})&\text{(for matter dominated universe)}\label{uniwave2}\end{cases}\e
where we have introduced the maximum radius of the universe $a_{IR}$, and $\Delta a_{WKB}$ represents the length of the violation the WKB approximation, namely, we can not use the WKB approximation in the $|a-a^{*}|<\Delta a_{WKB}$ region. In (\ref{uniwave2}), the first term comes from the neighborhood of the critical point $a^{*}$, and the second term comes form the IR region. In this approximation, the both results of  (\ref{uniwave2}) have a peak at the point on the critical surface where
\be R(\lambda_{i})\h{1mm}(M(\lambda_{i})) \h{2mm}\text{becomes maximum for the radiation (matter) dominated universe}.\nonumber\e
Therefore, the probability density (\ref{pro2}) has a strong peak on the critical surface where $R(M)$ is maximized, which we call the maximum entropy (matter) principle. 

\def\thesection{Appendix \Alph{section}}
\section{Sphaleron Process and its $v_{h}$ Dependence}\label{app:sphaleron}
In this appendix, we use the convention in \cite{Buchmuller:2005eh}. In the SM, the baryon and lepton number are not conserved as a consequence of the anomaly. This is related to the topological charge of the SU(2) gauge field,
\be B(t_{f})-B(t_{i})=N_{f}\{N_{cs}(t_{f})-N_{cs}(t_{i})\},\e
where 
\be N_{cs}(t)=\frac{g^{2}}{32\pi^{2}}\int d^{3}x\epsilon^{ijk}\text{Tr}[A_{i}\partial_{j}A_{k}+\frac{2}{3}igA_{i}A_{j}A_{k}]\e
is the Cern-Simon number, and $N_{f}$ is the number of the generations. The transition rate at zero temperature is given by the instanton action,
\begin{align} \Gamma&\simeq e^{-S_{\text{Ins}}}=e^{-\frac{4\pi}{g^{2}}}\nonumber\\
&={\cal{O}}(10^{-165}).
\end{align}  
Since this rate is very small, the violation of B (or L) does not occur at zero temperature. However, in a thermal bath, one can make transition between the gauge vacua, through thermal fluctuations. This transition rate is determined by the sphaleron configuration, a classical solution of the field theory \cite{Klinkhamer:1984di}. The transition rate of this process is \cite{Buchmuller:2005eh}
\be \Gamma_{sph}\simeq\alpha_{W}^{4}\times T\times e^{-\frac{E_{sph}}{T}}\e
where $\alpha_{W}=g_{2}^{2}/4\pi$ and $E_{sph}$ is the sphaleron energy. If the expansion rate of the universe $\frac{\dot{a}}{a}=H$ becomes lager than $\Gamma_{aph}$, the sphaleron process does not occur, and the total baryon number is fixed. The decoupling temperature is determined by equating $H$ and $\Gamma_{sph}$;
\be H\simeq\frac{T_{dec}^{2}}{M_{pl}}\simeq\alpha_{W}^{4}\times T_{dec}\times e^{-\frac{E_{sph}}{T_{dec}}}.\label{sphalerondec}\e
Note that we have used $H\simeq\frac{T^{2}}{M_{pl}}$ because this process happens in the radiation dominated era. The important fact is that, if $T_{dec}$ is comparable with the mass of the heavy particle, its abundance is suppressed by the factor $e^{-\frac{m}{T_{dec}}}$. 
To analyze the decoupling temperature, we will use the recent numerical result \cite{D'Onofrio:2012ni} in which the gauge couplings and the Higgs self couplings are fixed to the experimental value. One can obtain the sphaleron energy from the Fig.2 presented in \cite{D'Onofrio:2012ni} as
\begin{align} \log\left(\frac{\Gamma_{sph}}{T}\right)&=\log\left(\alpha_{W}^{4}\right)-\frac{E_{sph}(T)}{T}\nonumber\\
&\simeq \log(10^{-6})+123\times\frac{T-T_{c}}{T},\end{align}
which leads to
\be \Gamma_{sph}\simeq T\times e^{-13.8+123\frac{T-T_{c}}{T}},\e
where $T_{c}=160$GeV is the critical temperature of the phase transition. By replacing $T_{c}$ to
\be T_{c}=160\text{GeV}\times\frac{v_{h}}{246\text{GeV}},\e
we obtain the $v_{h}$ dependence of these quantities when the gauge couplings and the Higgs self coupling are fixed to the experimental values.  Then, solving Eq.(\ref{sphalerondec}), we obtain the decoupling temperature
\be T_{dec}\simeq\frac{7}{8}T_{c}=\frac{7}{8}\times160\text{GeV}\times\frac{v_{h}}{246\text{GeV}}=140\text{GeV}\times\frac{v_{h}}{246\text{GeV}}.\label{sphdec}\e

The number density of the particle species i is given by
\begin{align} n_{i}&=\frac{g_{i}}{(2\pi)^{3}}\int\frac{dp^{3}}{\exp\left(\frac{\sqrt{p^{2}+m_{i}^{2}}-\mu_{i}}{T}\right)\pm1}\nonumber\\
&=4\pi g_{i}\left(\frac{T}{2\pi}\right)^{3}\int_{0}^{\infty}dx\frac{x^{2}}{\exp\left(\sqrt{x^{2}+(\frac{
m_{i}}{T})^{2}}-\frac{\mu_{i}}{T}\right)\pm1}\label{density1}\end{align}
where $g_{i}$ represents the degree of freedom, and the sign $\pm$ is $-$ for bosons and $+$ for fermions. We can obtain the antiparticle density $\bar{n}_{i}$ by replacing $\mu_{i}$ with $-\mu_{i}$ in Eq.(\ref{density1}). Then, the difference between the number density of particle and antiparticle is
\begin{align} n_{i}-\bar{n}_{i}&=4\pi g_{i}\left(\frac{T}{2\pi}\right)^{3}\int_{0}^{\infty}dx\h{1mm}x^{2}\times\left(\frac{1}{\exp\left(\sqrt{x^{2}+(\frac{
m_{i}}{T})^{2}}-\frac{\mu_{i}}{T}\right)\pm1}-\frac{1}{\exp\left(\sqrt{x^{2}+(\frac{
m_{i}}{T})^{2}}+\frac{\mu_{i}}{T}\right)\pm1}\right)\nonumber\\
&\simeq8\pi g_{i}\mu_{i}\frac{T^{2}}{(2\pi)^{3}}\int_{0}^{\infty}dx\h{1mm}x^{2}\times\frac{\exp\sqrt{x^{2}+(m_{i}/T)^{2}}}{\left(\exp\sqrt{x^{2}+(m_{i}/T)^{2}}\pm1\right)^{2}}\nonumber\\
&=2\pi g_{i}\mu_{i}\frac{T^{2}}{(2\pi)^{3}}\int_{0}^{\infty}dx\h{1mm}x^{2}\begin{cases}\frac{1}{\cosh^{2}(\frac{\sqrt{x^{2}+(m_{i}/T)^{2}}}{2})}&\text{(for fermion)}\\
\\
\frac{1}{\sinh^{2}(\frac{\sqrt{x^{2}+(m_{i}/T)^{2}}}{2})}&\text{(for boson)}\end{cases}\nonumber\\
&:=g_{i}\mu_{i}\frac{T^{2}}{6}\begin{cases}g_{f}(m_{i})&\text{(for fermion)},\\
\\
g_{b}(m_{i})&\text{(for boson)},\end{cases}\end{align}
where
\be g_{f}(m_{i}):=\frac{3}{2\pi^{2}}\int_{0}^{\infty}dxx^{2}\frac{1}{\cosh^{2}(\frac{\sqrt{x^{2}+(m_{i}/T)^{2}}}{2})},\nonumber\e
\be g_{b}(m_{i}):=\frac{3}{2\pi^{2}}\int_{0}^{\infty}dxx^{2}\frac{1}{\sinh^{2}(\frac{\sqrt{x^{2}+(m_{i}/T)^{2}}}{2})}.\label{integral1}\e
Here we can take $\frac{T^{2}}{6}$ as a unit:
\be n_{i}-\bar{n}_{i}=g_{i}\mu_{i}\begin{cases}g_{f}(m_{i})&\text{(for fermion)}\\
\\
g_{b}(m_{i})&\text{(for boson)}.\end{cases}\e
where $g_{f}(0)=1,g_{b}(0)=2$. Note that the above density difference depends linearly on the chemical potential $\mu_{i}$. We denote the chemical potentials for the conserved quantities as $\mu_{a}$. Since the chemical potentials $\{\mu_{i}\}$ are conserved in thermal equilibrium, they can be written as linear combinations of the conserved quantum numbers;
\be \mu_{i}=\sum_{a}q_{ia}\mu_{a}.\e
In the broken phase, since the conserved quantum numbers are $B-L$ and the electromagnetic charge $Q$, we can write the chemical potentials $\mu_{i}$ as follows:
\be \mu_{u}=\frac{1}{3}\mu_{B-L}+\frac{2}{3}\mu_{Q}\e
\be \mu_{d}=\frac{1}{3}\mu_{B-L}-\frac{1}{3}\mu_{Q}\e
\be \mu_{e}=-\mu_{B-L}-\mu_{Q}\e
\be \mu_{\nu_{e}}=-\mu_{B-L}\e
\be \mu_{W}=-\mu_{Q}.\e
We can eliminate $\mu_{Q}$ by using the fact that $Q$ is zero:
\begin{align} Q=&\frac{2}{3}\sum_{i:\text{u sector}}g_{i}g_{f}(m_{i})\left(\frac{1}{3}\mu_{B-L}+\frac{2}{3}\mu_{Q}\right)-\frac{1}{3}\sum_{i:\text{d sector}}g_{i}g_{f}(m_{i})\left(\frac{1}{3}\mu_{B-L}-\frac{1}{3}\mu_{Q}\right)\nonumber\\
&-\sum_{i:\text{e sector}}g_{i}g_{f}(m_{i})(-\mu_{B-L}-\mu_{Q})+3g_{b}(m_{W})\mu_{Q}=0.\end{align}
This leads to
\be \mu_{Q}=-\mu_{B-L}\frac{4\sum_{i:\text{u}}g_{f}(m_{i})-2\sum_{i:\text{d}}g_{f}(m_{i})+6\sum_{i:\text{e}}g_{f}(m_{i})}{8\sum_{i:\text{u}}g_{f}(m_{i})+2\sum_{i:\text{d}}g_{f}(m_{i})+6\sum_{i:\text{e}}g_{f}(m_{i})+9g_{b}(m_{W})}.\e
As a result, the baryon and lepton number densities are given by
\begin{align} B&:=n_{B}-\bar{n}_{B}=6\cdot\frac{1}{3}\left(\frac{1}{3}\mu_{B-L}+\frac{2}{3}\mu_{Q}\right)\cdot\sum_{i:\text{u sector}}g_{f}(m_{i})+6\cdot\frac{1}{3}\left(\frac{1}{3}\mu_{B-L}-\frac{1}{3}\mu_{Q}\right)\cdot\sum_{i:\text{d sector}}g_{f}(m_{i})\nonumber\\
&=\frac{2\mu_{B-L}}{3}[\frac{6\sum_{i:\text{d}}g_{f}(m_{i})-6\sum_{i:\text{e}}g_{f}(m_{i})+9g_{b}(m_{W})}{8\sum_{i:\text{u}}g_{f}(m_{i})+2\sum_{i:\text{d}}g_{f}(m_{i})+6\sum_{i:\text{e}}g_{f}(m_{i})+9g_{b}(m_{W})}\sum_{i:\text{u}}g_{f}(m_{i})\nonumber
 \end{align}
\be\h{5cm}+\frac{12\sum_{i:\text{u}}g_{f}(m_{i})+12\sum_{i:\text{e}}g_{f}(m_{i})+9g_{b}(m_{W})}{8\sum_{i:\text{u}}g_{f}(m_{i})+2\sum_{i:\text{d}}g_{f}(m_{i})+6\sum_{i:\text{e}}g_{f}(m_{i})+9g_{b}(m_{W})}\sum_{i:\text{d}}g_{f}(m_{i})]\e
\begin{align} L&:=n_{L}-\bar{n}_{L}=2\cdot(-\mu_{B-L}-\mu_{Q})\sum_{i:e}g_{f}(m_{i})-3\mu_{B-L}\nonumber\\
&=\mu_{B-L}\{-2\frac{4\{\sum_{i:\text{u}}g_{f}(m_{i})+\sum_{i:\text{d}}g_{f}(m_{i})\}+9g_{b}(m_{W})}{8\sum_{i:\text{u}}g_{f}(m_{i})+2\sum_{i:\text{d}}g_{f}(m_{i})+6\sum_{i:\text{e}}g_{f}(m_{i})+9g_{b}(m_{W})}\sum_{i:\text{e}}g_{f}(m_{i})-3\}\end{align}
where we have assumed that neutrinos are massless particles. Once $N_{B-L}$ is given, the baryon number is given by
\be N_{B}=\frac{B}{B-L}N_{B-L}.\label{baryondef}\e
Since the general formula is very complicated, we do not write it here.

What we have to consider is whether the SM particles are relativistic or not at the  sphaleron decoupling temperature. If a particle is non-relativistic, the integral (\ref{integral1}) becomes very small, so it does not contribute to the baryon number. We can also read the $T$ dependence of the vacuum expectation value $v_{h}$ from the numerical result \cite{D'Onofrio:2012ni}:
\be v_{h}(T)=v_{h}\sqrt{\frac{T_{c}-T}{T_{c}}}.\e
Therefore, at the decoupling temperature, the quark mass is given by
\be  m_{i}(T_{dec})=\frac{y_{i}v_{h}}{\sqrt{2}}\sqrt{\frac{T_{c}-T_{dec}}{T_{c}}}=m_{i0}\sqrt{\frac{T_{c}-T_{dec}}{T_{c}}}\e
where $m_{i0}$ is the current quark mass, and $T_{dec}$ is given by Eq.(\ref{sphdec}). Since the suppression factors of the light particles start to have an effect at a very low temperature, it is almost valid to assume that all the particles except for the top quark and the W boson are massless. In this approximation, Eq.(\ref{baryondef}) becomes
\be N_{B}=\frac{2\{4+g_{b}(m_{W})\}\{5+g_{f}(m_{t})\}}{24\{5+g_{f}(m_{t})\}+g_{b}(m_{W})\{37+2g_{f}(m_{t})\}}N_{B-L}\label{baryon number}\e
This is the baryon number of the universe produced by the sphaleron process. We will use Eq.(\ref{baryon number}) in Section\ref{sec:baryon history} to estimate the radiation of the universe.

\def\thesection{Appendix \Alph{section}}
\section{Rough Estimations of $\tau_{He}$ and $\epsilon$}\label{app:epsilon}
\begin{figure}[h]
\begin{center}
\includegraphics[width=6cm]{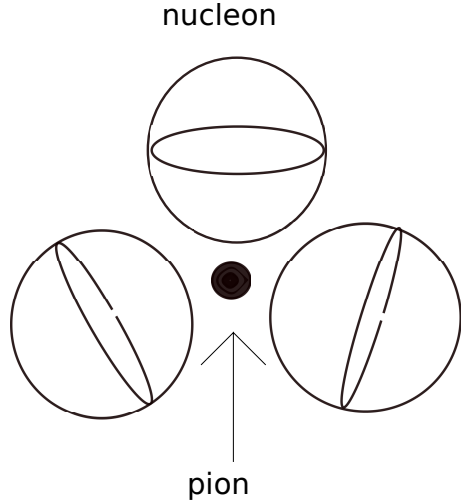}
\caption{The distribution of the remaining nucleons around the pion produced by the decay of a nucleon.}
\label{decay2}
\end{center}
\end{figure}
If we can apply the proton decay formula (\ref{eq:proton life}) to a nucleon in a helium nucleus, its life time is given by
\begin{align}\tau_{\text{N}}&\propto g^{-2}\frac{M_{X}^{4}}{(m_{p,n}-\delta)^{5}}(1-\frac{m_{\pi}^{2}}{(m_{p,n}-\delta)^{2}})^{-2}\nonumber\\
&\simeq\tau_{p}(1+5\frac{\delta}{m_{p,n}})\end{align}
where $\delta$ represents the fraction of the binding energy, and is expected to be 
\be 0<\delta<\Delta.\e
Here $\Delta=28$MeV is the binding energy of a helium nucleus. The life time of a helium nucleus is given by
\be \tau_{He}=\frac{\tau_{N}}{4}=\tau_{p}\frac{1+5\frac{\delta}{m_{p}}}{4},\e
because it has four nucleons. As a result, $\tau_{He}$ is in the region 
\be \frac{1}{4}\tau_{p}<\tau_{He}<\frac{7}{24}\tau_{p}.\label{maxhel}\e
In this paper, we sweep $\tau_{He}$ in this region. 

Another important fact about the atomic nuclei is that the pion produced by the decay of a nucleon in a helium nucleus may be scattered by the remaining nucleons, and loses its energy. See Fig.\ref{decay2}. We express this effect by the  dimensionless parameter $\epsilon$. Namely, $\epsilon$ represents the amount of the decrease of the radiation for each neutron due to the pion scattering. In principle, $\epsilon$ can be determined by the QCD, but here we give a rough estimate. The cross section of the pion-nucleon scattering is given by \cite{Beringer:1900zz}
\be \sigma\simeq3.0\text{fm}^{2},\e
and the radius of nucleon is
\be r_{N}=0.86\text{fm}.\e 
We assume that the pion produced by the decay of a nucleon in a helium nucleus is surrounded by the other nucleons with a distance
\be R\simeq2r_{N},\e
Then, the scattering probability is estimated as
\begin{align} P_{\text{sca}}&=\frac{3\sigma}{4\pi R^{2}}\nonumber\\
&\simeq\frac{1}{4}.\end{align}
When a free proton decays, it produces the radiation energy $m_{p}$, on the other hands, when a nucleon in a helium nucleus decays, the radiation is reduced to $\frac{1}{2}m_{p}$ if the pion loses the energy completely by the scattering. Therefore, the radiation energy is maximally reduced by $\frac{1}{4}m_{p}$ per neutrons. This means the maximum value of $\epsilon$ is 
\be \epsilon_{Max}=\frac{1}{4}\times P_{\text{sca}}=\frac{1}{16}.\label{epsilonmax}\e

\begin{figure}
\begin{center}
\includegraphics[width=8cm]{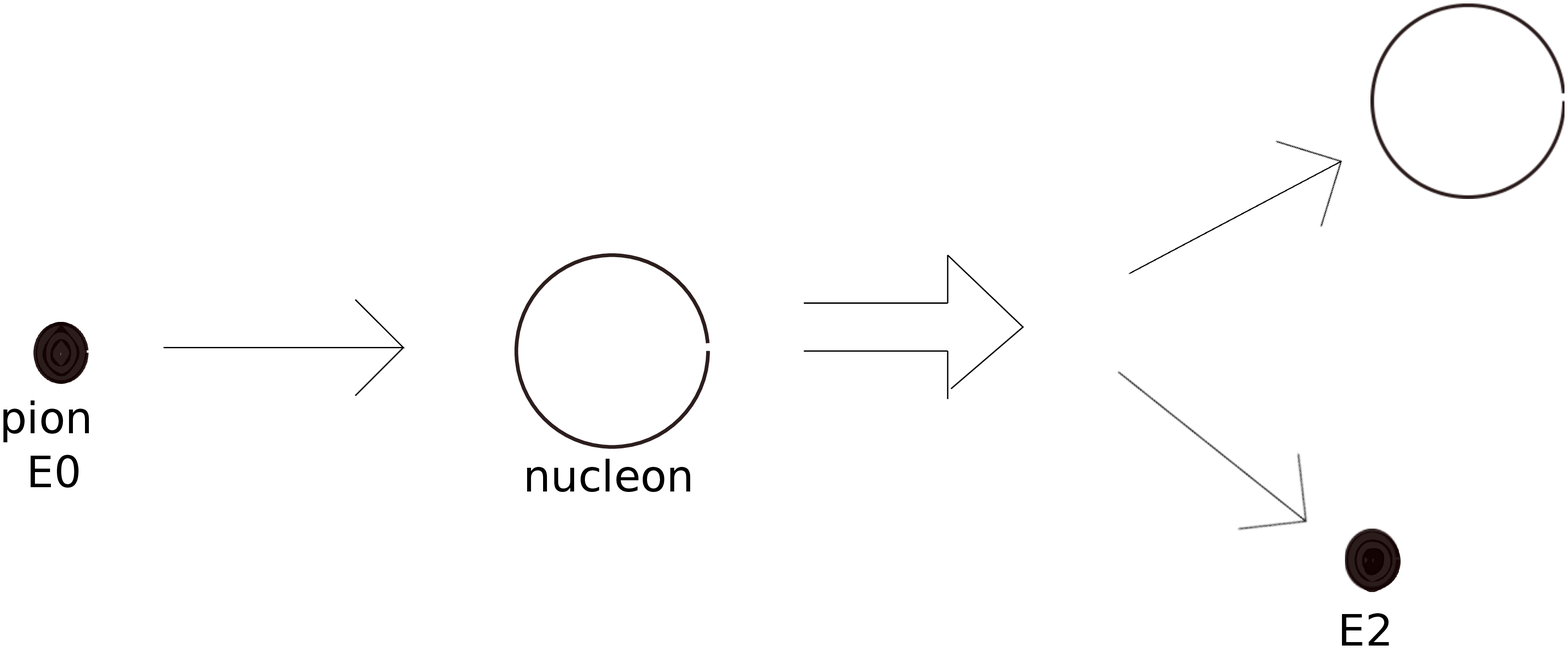}
\caption{}
\label{decay3}
\caption{The kinematic which shows that a pion is scattered by a nucleon and loses its energy partly.}
\end{center}
\end{figure}
\noindent The real value of $\epsilon$ is, of course, lower than Eq.(\ref{epsilonmax}), since the pion does not necessary lose all of its energy by the scattering. Fig.\ref{decay3} shows the kinematical situation. Since this is easily determined by solving the energy and momentum conservations, we do not follow the calculation here. When the pion is scattered to the angle $\theta=\pi$, the energy loss becomes maximum,
\be \frac{E_{2}}{E_{0}}(\theta=\pi)\simeq\frac{1}{2}.\e
Therefore, the real value of $\epsilon$ should be smaller than
\be \frac{1}{16}\times\frac{1}{2}=\frac{1}{32}.\e
To make more quantitative argument, we need to know how the nucleons are distributed in a helium nucleus.

\section{What if DM Dominates at the Late Stage of the Universe?}\label{app:darkmatter}
In this Appendix, we propose a scenario in which the DM physics is related to the Big Fix of $v_{h}$. As a simple extension of the SM which includes DM, we consider the Higgs portal scalar DM~\cite{DM,crosssection}. This DM only interacts with the Higgs field and its stability is guaranteed by $Z_2$ symmetry.
The Lagrangian is as follows:
\al{
\mathcal{L}=\mathcal{L}_{\text{SM}}
+\frac{1}{2}(\partial_{\mu}S)^2-\frac{1}{2}m_S^2S^2
-\frac{\rho}{4!}S^4-\frac{\kappa}{2}S^2 H^\dagger H,
}
where $S$ is the DM field.

Let us assume that, at the late stage of the universe, the energy of the universe is dominated by the DM. If this is the case, various parameters are fixed so that the energy of the DM, $M_\text{DM}=m_\text{DM} n_\text{DM} a^3_\text{end}$, is maximized.
This quantity is determined when the DM decouples from the thermal bath.
Following the argument of Lee and Weinberg~\cite{Lee:1977ua}, $M_\text{DM}$ is roughly proportional to
\al{
M_\text{DM}\propto\frac{1}{\langle \sigma v\rangle}
}
Here $\sigma$ and $v$ denote the annihilation cross section and the relative velocity, respectively. $\langle...\rangle$ means the thermal average.
\begin{figure}
\begin{center}
\hfill
\includegraphics[width=.6\textwidth]{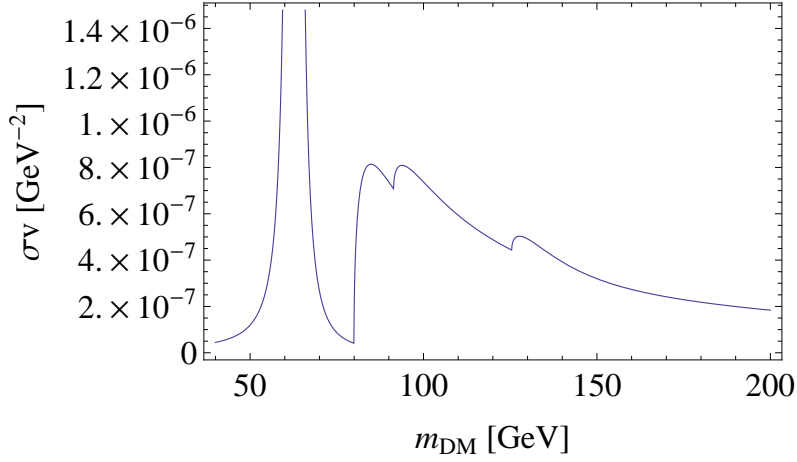}
\hfill\mbox{}
\caption{
The annihilation cross section of Higgs portal DM.
The parameters are set to be $\kappa=0.5, m_h=126\GeV$.
}
\label{crosssection_plot}
\end{center}
\end{figure}
The cross section of annihilation processes of the scalar singlet DM can be found in \cite{crosssection}:\footnote{This expression does not include the contributions from 4-body decay through virtual W and Z. Although such contributions become important around $m_S=m_{Z,W}$, this does not change our conclusion, and we ignore this effect.}
\al{\label{tree}
\langle\sigma v\rangle=&\frac{\kappa^2}{8\pi m_S^2\left[\left(4-\dfrac{m_h^2}{m_S^2}\right)^2+\dfrac{m_h^2}{m_S^2}\dfrac{\Gamma_h^2}{m_S^2}\right]}\nn
                                         =&
                                          \left\{
\begin{array}{ll}
6\dfrac{m_b^2}{m_S^2}\left(1-\dfrac{m_b^2}{m_S^2}\right)^{3/2}  & \quad b\bar{b}  \\ \\
\delta_{W(Z)}\left(\dfrac{m_{W(Z)}^2}{m_S^2}\right)^2
\left(2+\left(1-2\dfrac{m_S^2}{m_{W(Z)}^2}\right)^2\right)\sqrt{1-\dfrac{m_{W(Z)}^2}{m_S^2}}  & \quad WW(ZZ) \\ \\
2\left(\dfrac{\kappa}{\lambda}\dfrac{1-\frac{1}{4}\frac{m_h^2}{m_S^2}}{1-2\frac{m_S^2}{m_h^2}}
+1+\dfrac{1}{2}\dfrac{m_h^2}{m_S^2}\right)^2
\sqrt{1-\dfrac{m_h^2}{m_S^2}}
& \quad hh
\end{array}\right.
,
}
where $\delta_W=1,\delta_Z=1/2$ and $\Gamma_h=4.07$MeV.
We plot $\langle\sigma v\rangle^{-1}$ as a function of $m_\text{DM}$ in Fig.\ref{crosssection_plot}.
It is found that there are three local minima at $m_\text{DM}\simeq80, 90,$ and $126\GeV$. The global minimum corresponds to runaway $m_\text{DM}$. These local minima come from the threshold of W, Z boson and Higgs boson. 
This observation means that if $m_\text{DM}$ is equal to $m_W,m_Z$ or $m_h$, then $M_\text{DM}$ is locally maximized. However, singlet scalar with $m_\text{DM}\simeq80,90$GeV has already excluded by direct detection~\cite{crosssection}. 
Therefore, only the possibility is that the mass of the Higgs and that of the DM are the same, $m_h=m_\text{DM}\simeq126$GeV.
This possibility is testable by the direct detection experiments~\cite{direct_detection}.

\end{document}